\documentclass[11pt]{article}

\usepackage[preprint]{acl}
\usepackage[most]{tcolorbox}
\usepackage{times}
\usepackage{latexsym}
\usepackage{booktabs}
\usepackage{pifont}
\usepackage{xcolor}
\usepackage{twemojis}
\usepackage{enumitem}
\usepackage[T1]{fontenc}
\newcommand{\shield}{{\texttwemoji{1f6e1}}}
\usepackage[utf8]{inputenc}
\usepackage[table]{xcolor}
\usepackage{microtype}
\usepackage{subcaption} 
\usepackage{inconsolata}

\usepackage{graphicx}

\title{\shield \textit{RealRoute}: Dynamic Query Routing System via \\ Retrieve-then-Verify Paradigm}

\author{
Jiahe Liu$^1$\quad
Qinkai Yu$^2$\quad 
Jingcheng Niu$^3$\quad 
Xi Zhu$^4$\quad \\
\textbf{Zirui He}$^5$\quad
\textbf{Zhen Xiang}$^6$\quad
\textbf{Fan Yang}$^7$\quad 
\textbf{Jinman Zhao}$^{3}$ \thanks{corresponding author, corresponding email is ~\url{jzhao@cs.toronto.edu}}\\
  $^1$Technical University of Denmark\;\;\;  $^2$University of Exeter\;\;\;  $^3$University of Toronto\;\;\; \\$^4$Rutgers University \;\;\; $^5$ NJIT\;\;\; $^6$ University of Georgia \;\;\; $^7$ Wake Forest University\;\;\;
}

\begin{document}
    \maketitle
\begin{abstract}
Despite the success of Retrieval-Augmented Generation (RAG) in grounding LLMs with external knowledge, its application over heterogeneous sources (e.g., private databases, global corpora, and APIs) remains a significant challenge. Existing approaches typically employ an \textbf{LLM-as-a-Router} to dispatch decomposed sub-queries to specific sources in a predictive manner. However, this "LLM-as-a-Router" strategy relies heavily on the semantic meaning of different data sources, often leading to routing errors when source boundaries are ambiguous. In this work, we introduce \textit{\textbf{RealRoute System}}, a framework that shifts the paradigm from predictive routing to a robust \textbf{Retrieve-then-Verify} mechanism. RealRoute ensures \textit{evidence completeness through parallel, source-agnostic retrieval, followed by a dynamic verifier that cross-checks the results and synthesizes a factually grounded answer}. Our demonstration allows users to visualize the real-time "re-routing" process and inspect the verification chain across multiple knowledge silos. Experiments show that RealRoute significantly outperforms predictive baselines in the multi-hop Rag reasoning task. The RealRoute system is released as an open-source toolkit with a user-friendly web interface. The code is available at the URL: ~\url{https://github.com/Joseph1951210/RealRoute}.
\end{abstract}

\begin{table*}[t]
\centering
\resizebox{\textwidth}{!}{%
\begin{tabular}{lcccc}
\toprule
\textbf{System} & \textbf{Multi-Hop} & \textbf{Multi-Source Retrieval} & \textbf{Cross-Domain} & \textbf{Knowledge Check}\\
\midrule
Naive RAG~\cite{lewis2020retrieval}        & \textcolor{red}{\ding{55}} & \textcolor{red}{\ding{55}} & \textcolor{red}{\ding{55}} & \textcolor{red}{\ding{55}} \\
HippoRAG~\cite{jimenez2024hipporag}      & \textcolor{green!60!black}{\ding{51}} & \textcolor{red}{\ding{55}} & \textcolor{red}{\ding{55}} & \textcolor{red}{\ding{55}} \\
GraphRag system~\cite{edge2024local}              & \textcolor{red}{\ding{55}} & \textcolor{green!60!black}{\ding{51}} & \textcolor{red}{\ding{55}} & \textcolor{red}{\ding{55}} \\
Deepsieve system~\cite{guo2025deepsieve}                  & \textcolor{green!60!black}{\ding{51}} & \textcolor{green!60!black}{\ding{51}} & \textcolor{red}{\ding{55}} & \textcolor{red}{\ding{55}} \\
\midrule
\textbf{Real Route}         & \textcolor{green!60!black}{\ding{51}} & \textcolor{green!60!black}{\ding{51}} & \textcolor{green!60!black}{\ding{51}}& \textcolor{green!60!black}{\ding{51}} \\
\bottomrule
\end{tabular}}
\caption{Comparison of Real Route with representative RAG systems across key system capabilities, including multi-hop reasoning, multi-source retrieval, cross-domain generalization, and built-in knowledge verification.}
\label{tab:abacus_comparison}
\end{table*}

\section{Introduction}
Retrieval-Augmented Generation (RAG) has become a practical paradigm for question answering and reasoning in settings \cite{guo2025deepseek, asai2023self,baek2025probing} where the required information is distributed across multiple knowledge sources, such as general-purpose corpora, domain repositories, and organization-internal collections. In multi-hop question answering, a complex query is commonly decomposed into a sequence of dependent sub-questions, where each step requires retrieving evidence that is both relevant to the current sub-goal and compatible with downstream variable substitution and answer fusion \cite{gao2025synergizing}. A central systems question, therefore, concerns retrieval control: \textit{how to allocate a limited evidence budget across sources for each decomposed sub-query?}

A widely adopted design follows an LLM-as-a-Router strategy \cite{vsleher2025guarded,jiang2025droc,li2026llmrouterbench,guo2025deepsieve, shicommands}. The LLM predicts (before retrieval) the most appropriate source for the current sub-query before retrieval, and then retrieves exclusively from that source. Representative hard-routing systems~\citep{guo2025deepsieve} are efficient when source boundaries are clearly separable. However, in realistic deployments, sources often exhibit topical overlap or are artificially partitioned (e.g., English Wikipedia shards). In such cases, an incorrect pre-retrieval routing decision prevents critical evidence from entering the context window, becoming a dominant failure mode \cite{oche2025systematic}. Although reflection-based retry can partially mitigate this, it introduces additional routing rounds and remains constrained by single-source commitment.
\begin{tcolorbox}[
    enhanced,
    colframe=black!70,
    colback=yellow!5,
    boxrule=1pt, arc=2mm,
]


\textbf{Core Insight:} \textit{Real Route} eliminates routing hallucinations by shifting from error-prone predictive selection to a robust parallel retrieval and factual verification workflow.
\end{tcolorbox}

To address the limitations of existing LLM-as-a-Router strategies in RAG system, we present \textbf{\textit{Real Route}}, a lightweight multi-source RAG system that rethinks routing as an evidence-level verification process rather than a pre-retrieval source commitment. Instead of selecting a single source for each sub-query, Real Route performs parallel retrieval across all candidate sources and aggregates top-ranked results into a unified evidence pool. A dynamic selection module then filters this pool under a fixed context budget using retrieval scores, Reciprocal Rank Fusion (RRF), or an LLM-based relevance judge to construct the final context for answer generation. By decoupling evidence discovery from source prediction, RealRoute mitigates catastrophic failures caused by ambiguous or overlapping source boundaries, and maintains bounded context size and computational cost. 

\autoref{tab:abacus_comparison} compares Real Route with representative RAG systems along four practical capabilities: multi-hop reasoning, multi-source retrieval, cross-domain generalization, and built-in knowledge verification. Existing approaches typically optimize for only a subset of these requirements. Naive RAG pipelines lack explicit mechanisms for multi-hop decomposition or source coordination, while hard-routing frameworks such as HippoRAG and DeepSieve enable multi-hop reasoning but still rely on single-source commitment, limiting robustness when evidence is distributed across overlapping repositories. Graph-based retrieval systems improve multi-source aggregation but do not explicitly address cross-domain routing ambiguity or factual consistency checks. To best of our knowledge, no prior system simultaneously satisfies all four criteria. Real Route is designed to fill this gap by jointly supporting parallel multi-source retrieval, cross-domain evidence aggregation, and post-retrieval verification under a unified budgeted framework, enabling reliable performance in realistic deployments where knowledge boundaries are uncertain and heterogeneous. We further demonstrate RealRoute on multi-hop and biomedical QA benchmarks, and release an open-source implementation with an interactive interface that visualizes how evidence from multiple sources is aggregated into a unified pool and verified before answer.


\begin{figure*}[t]
  \centering
  \includegraphics[width=\textwidth]{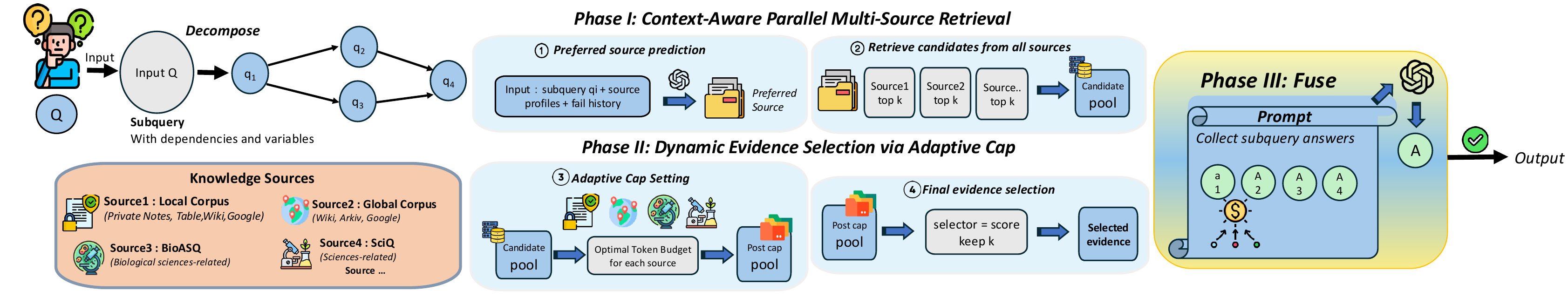}
\caption{\textbf{Overall framework of the multi-source retrieval and evidence check pipeline.} Phase I performs context-aware parallel retrieval by decomposing the input query into dependency-aware subqueries, predicting preferred sources, and retrieving top-k candidates from heterogeneous knowledge bases (e.g., local corpus, global corpus, BioASQ, SciQ..) to form a unified candidate pool. Phase II applies dynamic evidence selection via an adaptive token budget (cap) allocated per source, followed by score-based filtering to retain the most informative evidence. Phase III fuses the selected subquery answers within a structured prompt to produce the final output.}
  \label{fig:workflow}
\end{figure*}
\section{Related Work}
\paragraph{Agentic RAG}
Synergized RAG tightly interleaves retrieval and reasoning in an iterative loop, where intermediate hypotheses steer subsequent retrieval and new evidence updates the reasoning trajectory~\citep{rat,ircot,zhang2024raft,he-etal-2024-retrieving}. This coupling is often implemented via explicit workflows that structure the search space, including chain-based routines~\citep{rat,ircot,zhang2024raft,he-etal-2024-retrieving, jin2025disentangling}, tree-based variants~\citep{10.1609/aaai.v39i25.34876,kim-etal-2023-tree,hu-etal-2025-mcts}, and graph-based methods for multi-hop aggregation or exploration control~\citep{guo-etal-2025-lightrag,sun2024thinkongraph,jin-etal-2024-graph}. Agent orchestration further scales this loop from a single decision-maker to coordinated teams: single-agent systems enable tool-augmented search via prompting patterns~\citep{yao2022react,li-etal-2025-search}, supervised tool-use~\citep{schick2023toolformer}, or reinforcement learning~\cite{jin2025searchr}; multi-agent designs distribute roles through decentralized collaboration~\citep{wang-etal-2024-rag}, agent-based reasoning~\citep{wu2024avatar}, or centralized controllers that plan and route sub-tasks~\citep{10.1145/3746027.3754761}.


\paragraph{Reasoning-enhanced Retrieval}
Reasoning-enhanced RAG injects multi-step reasoning structure into retrieval~\citep{xu2025collabrag,jiang2025deepretrieval}. Collab-RAG emphasizes collaborative multi-perspective rewriting~\citep{xu2025collabrag}, while DeepRetrieval uses explicit decomposition signals to target intermediate premises~\citep{jiang2025deepretrieval}. Beyond rewriting, retrieval is increasingly planned and structured via multi-step planning~\citep{wang-etal-2024-learning-plan}, evidence fusion~\citep{chu-etal-2024-beamaggr,cheng-etal-2025-dualrag}, relevance filtering~\citep{zhao-etal-2024-seer,yoran2023making}, and structural inductive biases for coherent multi-hop retrieval~\citep{mavromatis-karypis-2025-gnn}. Since generation can still drift under noisy contexts, recent work studies context-aware utilization~\citep{islam-etal-2024-open}, evidence selection/reweighting~\citep{tran-etal-2025-rare}, and self-guided evidence-consistent reasoning~\citep{10.1609/aaai.v39i24.34743}, alongside traceability via post-hoc revision~\citep{gao-etal-2023-rarr} and explicit citation/verification~\citep{fang-etal-2024-trace}.

\paragraph{LLM-as-a-Router}
Multi-source RAG is increasingly cast as a routing problem. An LLM predicts which source/tool to invoke for each (sub-) query before retrieval. Search agents such as WebGPT~\citep{nakano2022webgptbrowserassistedquestionansweringhuman} treats information acquisition as sequential action selection. Recent work makes routing explicit in heterogeneous settings. Probing-RAG~\citep{baek-etal-2025-probing, jin2025exploring} leverages LLM self-reflection signals to guide document retrieval. OmniRouter~\citep{10.1145/3787470.3787480} performs cost-aware routing across multiple sub-indices that share a largely homogeneous retrieval interface. DeepSieve~\citep{guo2025deepsieve} performs sub-question-level knowledge routing, R1-Router~\citep{zhang2025routerr} dispatches across multiple knowledge bases, and dedicated query-routing frameworks assign queries to specialized retrieval tools without probing all sources~\citep{mu-etal-2024-query}. Related directions study source pruning/selection to reduce noise~\citep{10.1145/3746252.3761340} and robustness under distribution shift or ambiguous boundaries~\citep{vsleher2025guarded}. Fast use a VLLM as a router to route the query~\cite{sunvisual}.

\section{System Workflow}
In this section, we detail the operational workflow of our proposed system. As illustrated in \autoref{fig:workflow}, the pipeline executes this through three core phases: context-aware parallel retrieval, dynamic evidence selection, and adaptive generation. These phases collectively implement the capabilities summarized in \autoref{tab:abacus_comparison}, namely robust multi-hop reasoning, parallel multi-source retrieval, cross-domain evidence integration, and built-in verification.
\subsection{Context-Aware Parallel Multi-Source Retrieval}
For multi-hop reasoning tasks, the system initially decomposes the input question into a structured plan of dependent sub-queries. Before executing retrieval for a given sub-query $q_i$, the system resolves historical dependencies via a variable binding step (substitute\_variables). It dynamically injects answers produced by preceding sub-queries into placeholders within $q_i$, ensuring the current query possesses accurate contextual semantics. Subsequently, the system executes a parallel multi-source retrieval operation (retrieve\_multi\_source). Rather than consulting a single repository, the retriever fetches up to $k$ candidates (top\_k\_per\_source) from every available data source (e.g., local knowledge bases and global corpora), merging these independent streams into a unified candidate pool. Compared to prior LLM-as-a-Router approaches like DeepSieve \cite{guo2025deepsieve}, our parallel retrieval mechanism offers a distinct robustness advantage. While existing methods force the LLM to make a "hard routing" decision by predicting a single target source before any text is retrieved, our approach fundamentally eliminates this inherent single-point failure by prioritizing a comprehensive, cross-source candidate pool. Consequently, we bypass the fragility of predictive routing, where an incorrect guess under ambiguous source boundaries often leads to catastrophic evidence omission.
\subsection{Dynamic Evidence Selection via Adaptive Cap}
Once the global candidate pool is constructed, we perform evidence selection with Adaptive Cap, a preference-conditioned budgeting mechanism that instantiates our routing paradigm. Concretely, the system first obtains a preferred source label from the router. Adaptive Cap then applies a source-aware quota to the candidate pool: candidates retrieved from the preferred source are capped at a higher limit, while candidates from each non-preferred source are restricted to a lower limit. Following this initial quota application, the surviving candidates are globally ranked to fit within the strict final context budget. To accommodate diverse deployment constraints, this final ranking utilizes one of three plug-and-play selectors:

\begin{itemize}[leftmargin=*]\setlength\itemsep{-0.3em}

\item \emph{Score-based:} efficiently ranks and truncates candidates globally based on their raw retrieval scores.

\item \emph{RRF (Reciprocal Rank Fusion):} mitigates score scale discrepancies across heterogeneous sources by fusing relative rankings.

\item \emph{LLM-as-a-Judge:} leverages an LLM to evaluate fine-grained relevance, combining this judgment with retrieval scores.
\end{itemize}
This elegant transformation via Adaptive Cap prevents any single source from monopolizing the context window. Even if the initial routing preference is suboptimal, the system safely bypasses predefined boundaries to feed the most valuable cross-source evidence into the context, maximizing recall while strictly preventing context bloat.
\subsection{Adaptive Generation and Reflection}
In the final phase, the system constructs a synthesis prompt by concatenating the contextualized sub-query $q_i$ with the optimized evidence subset. The generator outputs a structured response containing the sub-query answer, a reasoning path grounded in the evidence, and a boolean evaluation flag. If the model determines the provided evidence is insufficient , it triggers an automated reflection loop. Each retry updates a failure context that discourages the system from repeating the same ineffective routing preference, prompting Adaptive Cap to explore alternative evidence allocation strategies. Once all sub-queries are successfully resolved, the system fuses the intermediate answers and traces into a comprehensive response for the original question.

\begin{algorithm}[t]
\caption{Routing-Conditioned Multi-Source Retrieval with Adaptive Cap}
\label{alg:adaptive-cap}
\begin{algorithmic}[1]
\Require Subquery $q$; source retrievers $\mathcal{R}=\{(s_1,r_1),\dots,(s_m,r_m)\}$; source profiles $\mathcal{P}$;
top-$K_s$ per source; final evidence budget $K$; preferred cap $C_{\text{pref}}$; other cap $C_{\text{other}}$; selector $\tau$
\Ensure Answer $a$; selected evidence $E$

\State $q_{\text{inst}} \gets \textsc{VariableBind}(q,\ \textit{previous\_answers})$
\State $s^\star \gets \texttt{None}$
\If{$(C_{\text{pref}}>0 \land C_{\text{other}}>0)$ \textbf{or} $\tau=\textsc{RoutingWeighted}$}
    \State $s^\star \gets \textsc{Route}(q_{\text{inst}},\mathcal{P})$
\EndIf

\State $C \gets \emptyset$
\ForAll{$(s,r)\in \mathcal{R}$}
    \State $D_s \gets \textsc{Retrieve}(r,\ q_{\text{inst}},\ K_s)$ \Comment{top-$K_s$ from source $s$}
    \State $C \gets C \cup \textsc{AnnotateSource}(D_s,\ s)$
\EndFor

\If{$C_{\text{pref}}>0 \land C_{\text{other}}>0 \land s^\star \neq \texttt{None}$}
    \State $C' \gets \textsc{ApplyAdaptiveCap}(C,\ s^\star,\ C_{\text{pref}},\ C_{\text{other}})$
\Else
    \State $C' \gets C$
\EndIf

\State $E \gets \textsc{SelectEvidence}(C',\ K,\ \tau)$ \Comment{e.g., score-based top-$K$}
\State $a \gets \textsc{GenerateAnswer}(q_{\text{inst}},\ E)$

\If{\textsc{ReflectionEnabled} \textbf{and} \textsc{Failed}$(a)$}
    \State \textsc{UpdateFailHistory}$(\dots)$
    \State \textbf{retry} up to \textsc{MaxReflexionTimes}
\EndIf

\State \Return $a,\ E$
\end{algorithmic}
\end{algorithm}

\section{Realworld Application}
The proposed system is a cross-domain, evidence-grounded research assistant for multi-hop question answering, designed to integrate organization-specific knowledge with external and specialized corpora. It supports flexible 2–4 source configurations and contrasts traditional Hard Routing (single-source retrieval) with Adaptive Cap, which preserves routing preference while enabling balanced multi-source evidence selection under a capped allocation strategy. The workflow includes query decomposition, variable binding, multi-source retrieval, preferred-source identification, capped evidence selection, answer generation, and optional reflection retries. The interface exposes routing decisions, retrieved evidence with scores, final and fallback answers, and quality/efficiency metrics (e.g., EM, F1, latency, token usage) for transparent inspection. Rather than claiming universal gains, the system contributes a practical, inspectable retrieval strategy that improves evidence coverage under routing uncertainty while remaining compatible with standard hard-routing pipelines.

\section{Result}
\begin{table*}[!ht]
\centering
\small
\setlength{\tabcolsep}{9pt}
\begin{tabular}{lllrrr}
\toprule
\textbf{Model} & \textbf{Setting} & \textbf{Method} & \textbf{EM} & \textbf{F1} & \textbf{Avg Tokens} \\
\midrule
GPT-4o & 2-source & Hard Routing (Baseline) & 49.30 & 61.70 & 3926.6 \\
\rowcolor{blue!5} 
 & & \textbf{Real Route (Ours)} & \textbf{53.67} & \textbf{66.93} & \textbf{2711.5} \\
\cmidrule{2-6}
 & 3-source & Hard Routing (Baseline) & 52.33 & 63.90 & \textbf{1881.6} \\
\rowcolor{blue!5}
 & & \textbf{Real Route (Ours)} & \textbf{53.00} & \textbf{64.61} & 2553.6 \\
\cmidrule{2-6}
 & 4-source & Hard Routing (Baseline) & \textbf{56.67} & \textbf{68.08} & \textbf{1941.2} \\
\rowcolor{blue!5}
 & & \textbf{Real Route (Ours)} & \textbf{56.67} & 67.09 & 2651.3 \\
\midrule
GPT-4o-mini & 2-source & Hard Routing (Baseline) & 46.09 & 56.89 & 2230.1 \\
\rowcolor{blue!5}
 & & \textbf{Real Route (Ours)} & \textbf{48.00} & \textbf{59.61} & 2785.6 \\
\cmidrule{2-6}

 & 3-source & Hard Routing (Baseline) & \textbf{46.00} & \textbf{58.57} & \textbf{2227.7} \\
\rowcolor{blue!5}
 & & \textbf{Real Route (Ours)} & 45.67 & 57.53 & 3051.21 \\
\cmidrule{2-6}

 & 4-source & Hard Routing (Baseline) & 48.67 & \textbf{60.63} & \textbf{2269.51} \\
\rowcolor{blue!5}
 & & \textbf{Real Route (Ours)} & \textbf{49.00} & 60.02 & 3183.29 \\
\midrule
\end{tabular}
\caption{Overall question answering performance. Hard Routing serves as the baseline. The best results within each setting are highlighted in \textbf{bold}, and our proposed method is emphasized with a shaded background.}
\label{tab:overall-metrics}
\end{table*}
This section investigates the question answering performance of Hard Routing and Real Route across diverse multi source environments. Furthermore, we conduct a paradigm comparison to benchmark our method against representative RAG baselines.

\subsection{Experimental Setup and Source Configurations}
\paragraph{Evaluation Protocol.} All results are reported on a deterministic evaluation subset constructed via evenly spaced index sampling. This removes randomness from query selection and ensures that different retrieval policies are evaluated on exactly the same queries within each dataset. We report Exact Match (EM) and token-level F1 computed on normalized answer strings. Normalization lowercases text, removes punctuation, collapses whitespace, and removes a fixed set of stopwords. EM is defined as exact equality after normalization, and F1 is computed based on normalized token overlap.

\paragraph{Source Configurations.}
We consider three multi-source configurations that vary in semantic proximity and the number of domain:

\begin{itemize}[leftmargin=*]\setlength\itemsep{-0.3em}

\item \emph{Two source setting (\citealp[hotpot\_qa]{yang-etal-2018-hotpotqa})}: evidence is partitioned into \texttt{local} and \texttt{global}, following the original DeepSieve split. 
 These two labels represent closely related retrieval pools rather than different domains. Both are Wikipedia derived corpora, with the distinction intended to support complementary retrieval behavior rather than to encode fundamentally different semantics.

\item \emph{Three source setting (multi\_source):} queries are labeled with one of \texttt{wiki}, \texttt{sciq}~\citep{welbl-etal-2017-crowdsourcing}, or \texttt{bioasq}~\citep{tsatsaronis2015overview}, corresponding to general encyclopedic knowledge, science exam–style knowledge, and biomedical knowledge, respectively.

\item \emph{Four source setting (mixed\_4source)}: queries are labeled as \texttt{hotpot}~\citep{{yang-etal-2018-hotpotqa}}, \texttt{sciq}~\citep{welbl-etal-2017-crowdsourcing}, or \texttt{bioasq}~\citep{tsatsaronis2015overview}. In this setting, hotpot aligns with Wikipedia style knowledge and is operationalized in the system through the two closely related retrieval pools \texttt{local} and \texttt{global}. And \texttt{sciq} and \texttt{bioasq} remain domain specific corpora.
\end{itemize}

Overall, these settings offer a spectrum of semantic proximity, ranging from the homogeneous Wikipedia-derived pools in the two-source setting to the highly heterogeneous domains in the three- and four-source settings.
\subsection{Overall Performance}
\autoref{tab:overall-metrics} summarizes the main results for GPT-4o and GPT-4o-mini, reporting average prompt tokens per query as a coarse proxy for inference cost. In the two source condition, Adaptive Cap improves both EM and F1 over Hard Routing, increasing EM from 49.30 to 53.67 and F1 from 61.70 to 66.93, while also reducing average prompt tokens per query from 3926.6 to 2711.5. This pattern is consistent with the intuition that, when the two sources are semantically close, Adaptive Cap can reduce redundant evidence collection and stabilize fusion with a smaller prompt budget.

Conversely, in highly heterogeneous multi-source scenarios (three- and four-source), a different consistent pattern emerges: Hard Routing is inherently more token-efficient, whereas Adaptive Cap addresses the varying utility of different knowledge sources by dynamically adjusting its evidence selection. This flexibility intentionally allocates additional tokens to support broader evidence coverage and more robust cross-source aggregation. While this incurs a higher inference cost, it directly benefits the final prediction without introducing volatility. Taking the three-source setting as an example, Real Route yields consistent improvements in accuracy (EM increasing to 53.00 and F1 to 64.61); the increased token usage (2553.6 vs. 1881.6) reflects the system's ability to aggregate more relevant context when necessary. As the source mixture further expands to the four-source condition, the fused-answer metrics remain highly stable (EM tied at 56.67, with Hard Routing maintaining a slightly higher F1 of 68.08 vs. 67.09 at a lower token cost). Together, these trends demonstrate that Real Route provides a dependable performance profile under heterogeneous source mixes, successfully trading a controlled extra budget for broader evidence coverage without compromising overall reliability.

\subsection{Paradigm Comparison}
\begin{table}[t]
\centering 
\resizebox{0.97\linewidth}{!}{%
\begin{tabular}{l|rrr}
\toprule
\textbf{Paradigm} & \textbf{EM} & \textbf{F1} & \textbf{\#Tokens} \\
\midrule
Direct & 36.2 & 28.0 & \textbf{55.5} \\
IR-CoT \cite{trivedi2023interleaving}& 30.8 & 22.4 & 481.9 \\
ReAct \cite{yao2022react}& 39.6 & 32.2 & 9795.1 \\
ReWOO \cite{xu2023rewoo}& 40.1 & 30.4 & 1986.2 \\
Reflexion \cite{shinn2023reflexion}& 46.7 & 62.5 & 37893.0 \\
DeepSieve \cite{guo2025deepsieve}& 49.3 & 61.7 & 3926.6 \\
\rowcolor{blue!5}
\textbf{Real Route} & \textbf{53.7} & \textbf{66.9} & 2711.6 \\
\bottomrule
\end{tabular}}
\caption{Comparison across QA reasoning paradigms on HotPotQA. 
All results are reported under the same benchmark configuration. The best results within each setting are highlighted in \textbf{bold}, and our proposed method is emphasized with a shaded background.}
\label{tab:reasoning-paradigm}
\end{table}

\autoref{tab:reasoning-paradigm} compares Real Route with representative QA paradigms. These methods differ in how they retrieve and use external evidence. Except for ``Real Route'', all baseline numbers come from DeepSieve under the same benchmark configuration (on HotPot QA~\cite{yang-etal-2018-hotpotqa}).

\paragraph{Paradigms.}
We compare RealRoute against representative QA paradigms that differ in how they retrieve and use external evidence.
Direct answers without retrieval. IR-CoT~\citep{trivedi2023interleaving} adds an explicit step-by-step reasoning trace but still uses no external evidence.
ReAct~\citep{yao2022react} alternates reasoning steps and retrieval actions. ReWOO~\cite{xu2023rewoo} separates planning from evidence gathering to reduce unnecessary tool calls. Reflexion~\citep{shinn2023reflexion} adds self-critique and retry, often producing longer trajectories. DeepSieve \cite{guo2025deepsieve} is the hard routing-based RAG baseline. It selects one source per sub-query before retrieval.

\paragraph{Findings.}
Purely generative methods such as Direct and IR-CoT use fewer tokens but perform poorly on multi-hop factual QA due to the lack of external evidence. Agentic reasoning methods (ReAct, ReWOO, Reflexion) improve accuracy. They also increase token usage because they extend the reasoning trajectory. DeepSieve achieves a better accuracy and cost balance.

RealRoute improves EM from 49.3 to 53.7 and F1 from 61.7 to 66.93 over DeepSieve. It also reduces token usage by about 31\%. This result supports our claim. Retrieve-then-select and adaptive evidence budgeting reduce routing errors and redundant context. They improve answer quality at lower generation cost.

\section{Conclusion}
We presented Real Route, a robust multi-source RAG framework that replaces error-prone predictive routing with a Retrieve-then-Verify paradigm. By combining parallel source-agnostic retrieval with Adaptive Cap for dynamic evidence budgeting, Real Route effectively mitigates routing hallucinations caused by ambiguous source boundaries. Evaluations across diverse multi-source settings demonstrate that Real Route consistently achieves superior accuracy-cost trade-offs compared to hard routing baselines and complex agentic paradigms. Real Route serves as a highly dependable and cost-efficient default for heterogeneous multi-source RAG applications.

\section*{Limitations}
Several important limitations of this study must be emphasized. RealRoute improves robustness to routing ambiguity through parallel retrieval and adaptive evidence budgeting, but it incurs higher retrieval overhead and relies on the quality of underlying retrievers and budget hyperparameters, without providing formal guarantees on optimal evidence allocation.

\bibliography{custom}
\clearpage
\appendix
\section{Appendix}
\subsection{Routing Prompt (Preferred Source Signal)}

The prompt shown below is used for source selection. 


\begin{lstlisting}[
language={},
frame=single,
breaklines=true,
breakatwhitespace=true,
columns=fullflexible,
basicstyle=\ttfamily\footnotesize
]
You are a routing assistant. Your task is to decide which knowledge source is most relevant for answering the given query.

Available knowledge sources:

{profiles_text}

QUERY:
{query}

{fail_history}

Please output ONLY the source name (one of: {choices_str}) that is most relevant to answer this query.
Do not add any explanation or extra words.
\end{lstlisting}

\label{sec:appendix}
\subsection{System Demo}
\begin{figure*}[t] 
    \centering
    \begin{subfigure}[b]{0.48\linewidth}
        \centering
        \includegraphics[width=\linewidth]{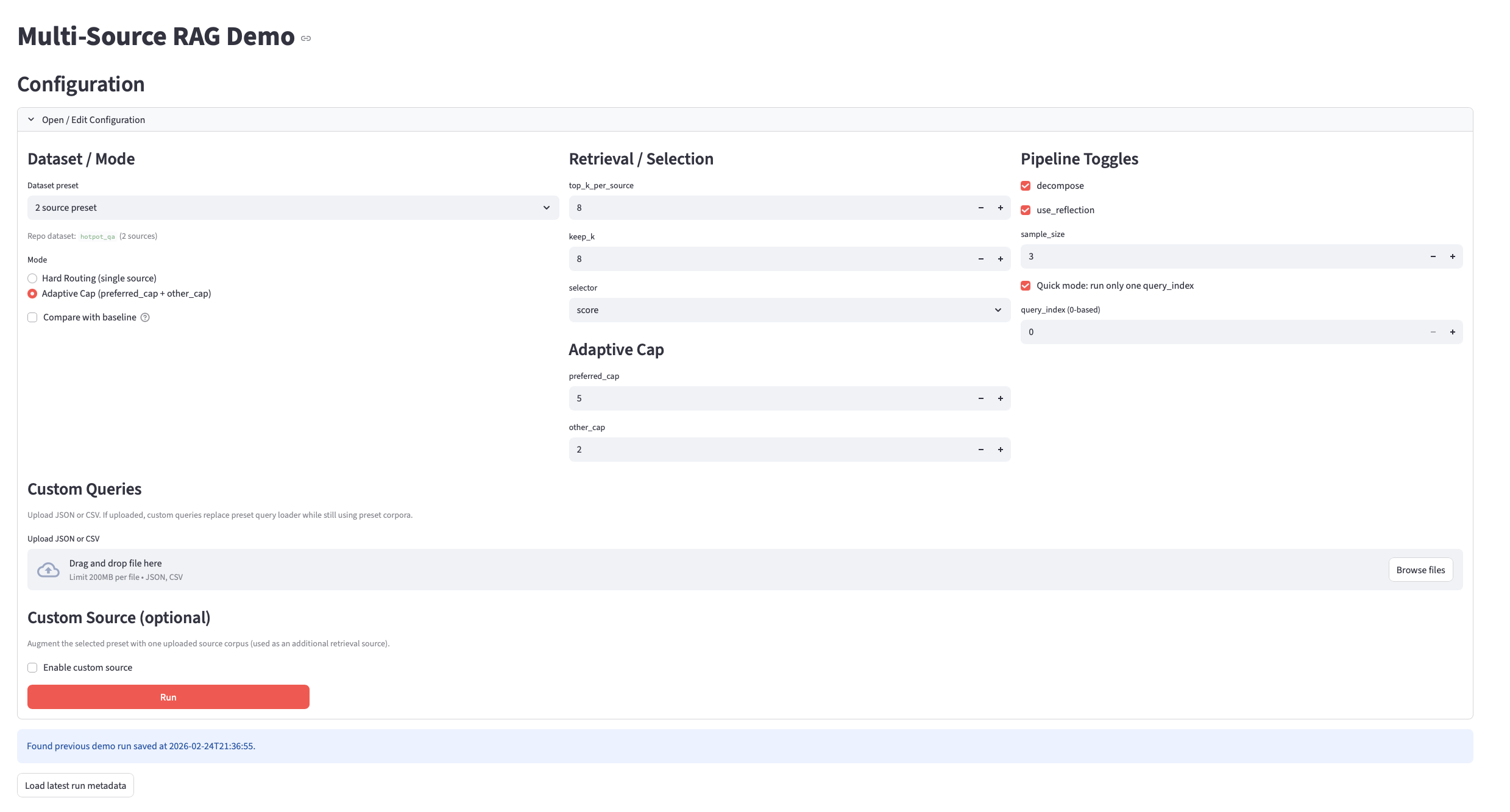}
        \caption{Configuration panel}
        \label{fig:demoUI1}
    \end{subfigure}
    \hfill 
    \begin{subfigure}[b]{0.48\linewidth}
        \centering
        \includegraphics[width=\linewidth]{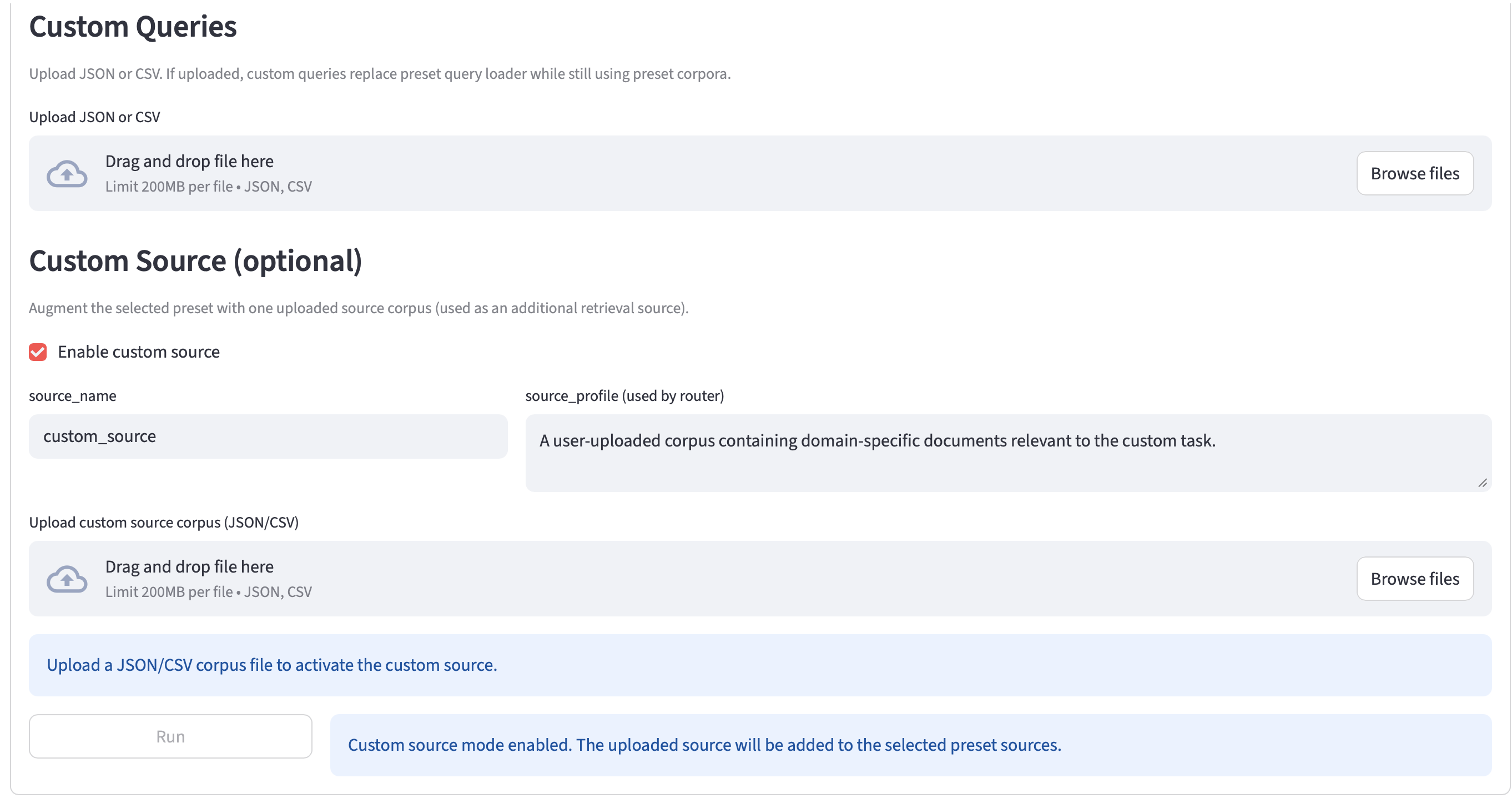}
        \caption{Custom Source panel}
        \label{fig:demoUI2}
    \end{subfigure}
    
    \caption{The user interface of the experiment: (a) shows the initial configuration, while (b) details the custom source settings.}
    \label{fig:demoUI_combined}
\end{figure*}
\autoref{fig:demoUI1} shows the experiment configuration panel of the demo, where the user can configure various parameters for the experiment, including dataset, mode, retrieval settings, and pipeline behaviors.

The user can select a dataset preset (e.g., "2 source preset") that determines the dataset used for the experiment (e.g., hotpot\_qa with 2 sources).

The user can choose between Hard Routing (single source) or Adaptive Cap. The configuration in the screenshot is set to Adaptive Cap, which uses parameters like \texttt{preferred\_cap} and \texttt{other\_cap}.

Users can configure the following retrieval parameters:
\begin{itemize}
    \item \texttt{top\_k\_per\_source}: Defines how many top candidates to retrieve from each source.
    \item \texttt{keep\_k}: Specifies how many candidates to retain after retrieval.
    \item \texttt{selector}: Specifies the method for selecting final evidence (e.g., score).
\end{itemize}

In Adaptive Cap mode, the user configures the values for \texttt{preferred\_cap} and \texttt{other\_cap} to control how evidence is capped based on the preferred source.

Users can toggle various settings for the pipeline, including:
\begin{itemize}
    \item \texttt{decompose}: Enables or disables decomposition of queries into subqueries.
    \item \texttt{use\_reflection}: Enables or disables reflection for retrying failed queries.
    \item \texttt{sample\_size}: Sets the sample size for evidence selection.
\end{itemize}


Users can choose to compare the current experiment with a baseline by running both configurations (Hard Routing and Adaptive Cap) on the same set of queries.

After selecting the desired parameters, users can click the \texttt{Run} button to trigger the backend pipeline with the selected configuration.

The \autoref{fig:demoUI2} illustrates the Custom Source Upload and Augmentation functionality within the demo interface. This feature enables users to augment the preset sources with custom datasets, which can be uploaded in JSON or CSV formats. The Custom Source Upload feature allows users to:
\begin{itemize}
    \item \textbf{Enable Custom Source}: Users can activate the custom source functionality by selecting the checkbox "Enable custom source".
    \item \textbf{Input Custom Source Details}: Users are prompted to provide a source name and a source profile. The source name will be used as an identifier for the uploaded data, and the source profile will provide metadata relevant for routing the queries to the appropriate source.
    \item \textbf{Upload Custom Corpus}: A user can upload a corpus in JSON or CSV format, which will be parsed into a set of documents for use in the retrieval process.
    \item \textbf{Run Button Activation}: The Run button remains disabled until a valid custom source is uploaded. Once the custom corpus is uploaded and parsed, the Run button becomes active, triggering the backend pipeline.
\end{itemize}

\begin{figure*} [h!]
    \centering
    \includegraphics[width=1.0\linewidth]{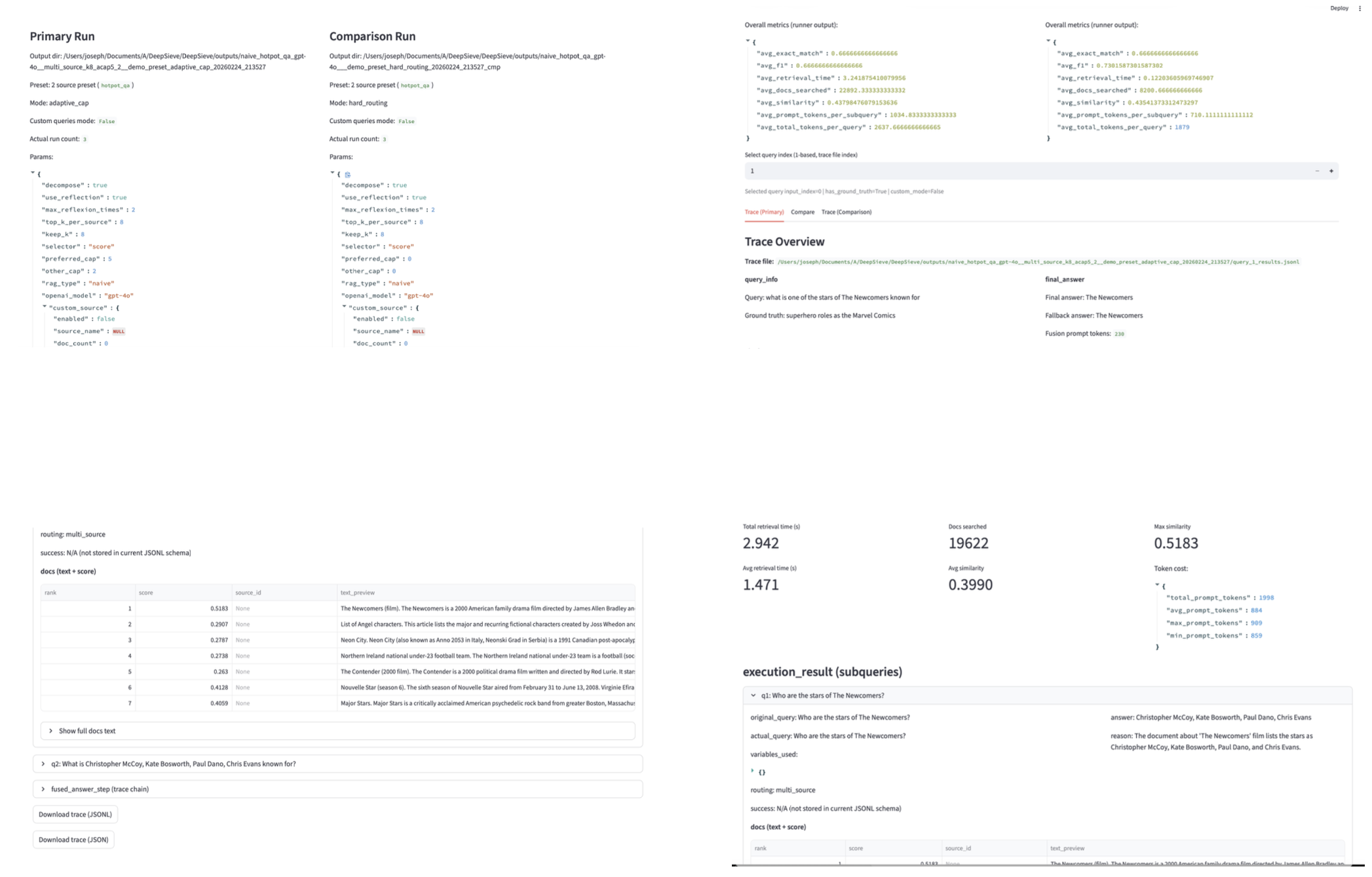}
    \caption{Comparison Run Panel}
    \label{fig:placeholder3}
\end{figure*}

\end{document}